\documentclass[reprint,aps,prl,showpacs]{revtex4-1}
\usepackage{graphicx,epsfig}
\usepackage{epstopdf}
\usepackage{bm}
\begin{document}

\title{Rotating quantum turbulence in superfluid $^4$He in the $T = 0$ limit}
\author{P. M. Walmsley and A. I. Golov}
\affiliation{School of Physics and Astronomy, The University of Manchester, Manchester M13 9PL, UK}
\date{\today}

\begin{abstract}
Observations of quantum turbulence in pure superfluid $^4$He in a rotating container are reported. New techniques of large-scale forcing (rotational oscillations of the cubic container) and detecting (monitoring ion transport along the axis of rotation) turbulence were implemented. Near the axial walls, with increasing forcing the vortex tangle grows without an observable threshold. This tangle gradually develops into bulk turbulence at a characteristic amplitude of forcing that depends on forcing frequency and rotation rate.  At higher amplitudes, the total vortex line length increases rapidly. Resonances of inertial waves are observed in both laminar and turbulent bulk states. On such resonances, the turbulence appears at smaller amplitudes of forcing. 
\end{abstract}


\maketitle\

In superfluid $^4$He driven to the state of {\it quantum turbulence} (QT), quantized vortex lines (each with circulation $\kappa = 1.00\times10^{-3}$\,cm$^2$/s) make a dynamic tangle \cite{Vinen}. At lengthscales smaller than the mean distance between the lines $\ell = L^{-1/2}$, where $L$ is their total length per unit volume, QT differs markedly from classical turbulence. In the zero-temperature limit $T=0$ viscosity vanishes, although finite dissipation \cite{WalmsleyPRL2007} is still maintained probably due to short-scale deformations of vortex lines (Kelvin waves) -- the cascade of energy to these lengthscales being maintained by line reconnections. 
One thus might need very small amplitudes of forcing to sustain steady turbulence. As the rate and efficiency of reconnections are expected to decrease with the alignment of neighbouring lines, to investigate the intrinsic mechanisms of dissipation of QT it is instructive to study the case of vortex tangles polarized by an imposed rotation. For instance, in classical fluids rotation modifies the dynamics of the energy cascade \cite{Lamriben2011} and generally opposes turbulence.

In rotating superfluid, the equilibrium state is that with an array of parallel vortex lines of a uniform density $L_0 = 2\Omega_0/\kappa$. Low-frequency ($\omega \leq 2\Omega_0$) excitations are expected to be inertial waves \cite{Hall1960} (in essence, oscillations of compression and bending of the vortex array), i.\,e. the same as in classical fluids \cite{Bewley2007,Lamriben2011}. Swanson {\it et al.} \cite{Swanson1983} studied rotating superfluid $^4$He at high temperatures, at which one can destabilize individual vortex lines via an axial counterflow of the normal and superfluid components. They observed two different turbulent states that appear after exceeding the critical values of the counterflow velocity. Using numerical simulations and disregarding vortex pinning at the container walls, Tsubota {\it et al.} \cite{Tsubota2004} interpreted the first transition as the proliferation of Kelvin waves on isolated rectilinear vortex lines (high degree of polarization, $L \approx L_0$) and the second one as the onset of a vortex tangle of low polarization ($L \gg L_0$) caused by reconnections of neighbouring vortex lines.

\begin{figure}
\centerline{\includegraphics[width=7.5cm]{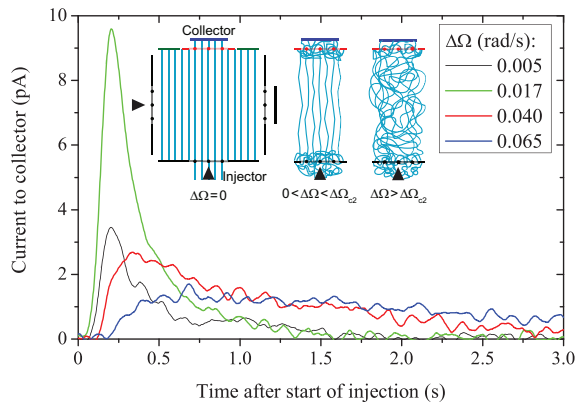}}
\caption{
(color online) Records of current to the top collector $I_{\rm coll}(t)$ after a 0.1\,s-long injection, beginning at $t=0$, at the bottom of the cell -- for different amplitudes of agitation $\Delta\Omega$. $\Omega_0 = 1.5$\,rad/s, $\omega = 1.5$\,rad/s, $T=0.2$\,K, $E=40$\,V/cm. Inset: sketch of a stationary cell with the vortex array (left); sketch of vortex lines in the axial region of the cell at moderate (center) and high (right) level of forcing. 
}
\label{FigPulsesForcedCell}
\end{figure}

The aim of this work was to observe and investigate QT in rotating superfluid $^4$He in the $T=0$ limit, driven at classical ($\gg \ell$) lengthscales; for instance, whether the critical amplitude of forcing is finite and if there could be more than one turbulent regime? The experiments were performed in isotopically-pure liquid $^4$He at temperature $T\leq 0.2$\,K and pressure 0.1\,bar. A cube-shaped volume of side $d=45$\,mm was confined by six square electrodes (see side view in inset in Fig.\,\ref{FigPulsesForcedCell}; details are in \cite{cell}). The whole cryostat could rotate around its vertical axis at a computer-controlled angular velocity $\Omega(t)$.  

To force turbulence, an AC component of frequency $\omega$ and small amplitude $\Delta \Omega$ was added to the DC angular velocity $\Omega_0$ of rotation of the cryostat, $\Omega(t) = \Omega_0 + \Delta\Omega {\rm sign}(\sin(\omega t))$. 
The following mechanisms of turbulence generation are expected: large-scale vortices are created after each stroke due to the flow separation near the corners of lateral walls; quantized vortices are agitated near the axial walls due to surface friction (pinning-unpinning); 
inertial waves induce large-scale AC flow in the container (in resonance conditions, its velocity can be much greater than $\sim d\Delta\Omega$). 

To detect turbulence in the middle of the container, we monitored the connectivity of vortex lines in the axial direction through measurements of the transport of ions trapped on them. 
 Negative ions (electrons in a bubble state) were injected from the field-emission tip 1\,mm below the centre of the grid in the bottom plate -- by applying voltage to the tip $V_{\rm tip} = 525$\,V (thus injecting the current of $\sim 100$\,pA that was mainly terminated at the injector grid) for 0.1\,s, and then reverting $V_{\rm tip}$ back to zero. At the opposite wall, the currents separately collected by the collector, $I_{\rm coll}(t)$, grid together with its frame, $I_{\rm grid}(t)$, and top plate, $I_{\rm plate}(t)$, were converted into voltage and recorded. The grid in front of the top collector, made of a square mesh of wire of diameter $w=0.02$\,mm with period $s=0.5$\,mm, was stretched on a metal ring-frame of 13\,mm i.\,d. and 19\,mm o.\,d. and was separated from the surrounding top plate by a 1\,mm gap -- thus making the radius of its gridded part $R_1=6.5$\,mm and effective radius dividing the grid and outer top-plate $R_2 \approx 10$\,mm. The time constant of the recording electronics was 30\,ms.
  All charges, currents and voltages are quoted with the opposite sign (i.\,e. assuming the electron's charge to be positive). The voltage between the top collector and its grid (and top plate), side plates, and bottom plate were  10\,V, 100\,V, and 190\,V, respectively -- making the mean axial field in the drift space $E=40$\,V/cm. 
  In Fig.\,\ref{FigPulsesForcedCell} we show $I_{\rm coll}(t)$, all at the same $\Omega_0 = 1.5$\,rad/s but for different amplitudes of forcing $\Delta\Omega$ at frequency $\omega = 1.5$\,rad/s. 
With increasing forcing, the amplitude $I^*$ of the peak of $I_{\rm coll}(t)$, firstly increases and then decreases. The peak's width at half-maximum, $\Delta t$, only begins to increase at high amplitudes of forcing. 
    
At $T \leq 0.2$\,K all negative ions in $^4$He are bound to vortex lines.
During the injection they create a compact charged vortex tangle near the tip that spreads through the bottom grid. 
With little forcing, there is an array of rectilinear vortex lines everywhere, except for this vortex tangle at the bottom that feeds ions into the array. The ions' time of flight along straight vortices to the top, at the terminal velocity of $\sim 10$\,m/s \cite{OstermeierGlaberson}, is just $\sim 4$\,ms. The position of the maximum at 0.20\,s and width at half-height $\Delta t = 0.21$\,s of $I_{\rm coll}(t)$ (Fig.\,\ref{FigPulsesForcedCell}) reflect the slower dynamics of the evolution of the charged vortex tangle near the injector and not the ion emission duration 0.10\,s: first ions cross the injector grid some 0.05\,s after the beginning of the injection, the maximal current is fed into rectilinear lines after further 0.15\,s,  
 followed by a nearly-exponential tail with the lifetime of some 0.25\,s.

 \begin{figure}
\centerline{\includegraphics[width=7.5cm]{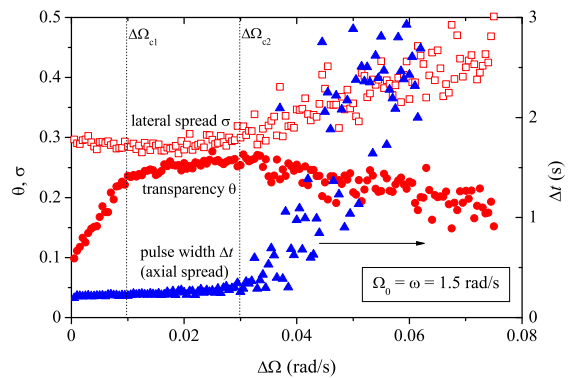}}
\caption{(color online)
Analysis of the records of currents to the top collector (Fig.\,\ref{FigPulsesForcedCell}), grid and surrounding plate for $\Omega_0 = \omega = 1.5$\,rad/s. Left axis: grid transparency $\theta$ (solid red circles), fraction of charge arriving outside the grid $\sigma$ (open red squares). Right axis: width $\Delta t$, of collector current pulses ($\triangle$). The vertical lines indicates the critical amplitudes of forcing $\Delta\Omega_{\rm c1} \approx 0.01$\,rad/s and $\Delta\Omega_{\rm c2} \approx 0.03$\,rad/s. 
}
\label{FigTranspSpreadWidth}
\end{figure}

The total charges arriving at different electrodes, $Q_{\rm coll}$, $Q_{\rm grid}$, $Q_{\rm plate}$, were obtained by integrating the corresponding currents. These allowed to quantify the fraction of ions arriving outside the radius of $R_2 \approx 10$\,mm (``lateral spread''), $\sigma = Q_{\rm plate}/(Q_{\rm coll}+Q_{\rm grid}+Q_{\rm plate})$, and transparency of the grid assembly, $\theta=Q_{\rm coll}/(Q_{\rm coll}+Q_{\rm grid})$. The lateral and axial spread parameters $\sigma$ and $\Delta t$ are indicative of bulk turbulence (at least when they grow above their no-forcing values, see below), while the grid transparency $\theta$ is a local measure of the state of vortex lines near the collector grid. When rectilinear vortex lines are not perturbed and their density $L_0$ is comparable to that of grid cells, $s^{-2}$, many lines terminate not at the collector but at the grid; hence, only a fraction of ions reach the collector. 
This fraction increases with increasing agitation of vortex lines that can cause intermittent reconections of lines between pinned (terminated on grid) and unpinned (transparent for ions) states.
When these reconnections result in a vortex tangle of density $L \gg s^{-2}$ near the grid, most ions that arrive within the gridded radius do make it to the collector, except for those heading to a grid's wire head-on within the catching diameter $w'=w+\gamma \ell$, where $\gamma \sim 1$. 
Thus, if ions arrive uniformly within a radius $R_2 \approx 10$\,mm or greater, 
\begin{equation}
	\theta \approx \left(\left(1-\frac{w'}{s}\right)\frac{R_1}{R_2}\right)^2 = 0.24,
\label{theta}
\end{equation} 
evaluated for $\gamma = 1$ and $L = \ell^{-2}= 10^4$\,cm$^{-2}$.  

The graphs of $\theta$, $\Delta t$ and $\sigma$ versus $\Delta\Omega$, corresponding to the records from Fig.\,\ref{FigPulsesForcedCell}, are shown in Fig.\,\ref{FigTranspSpreadWidth}.
There are two limiting regimes and their cross-over between the `critical' amplitudes $\Delta\Omega_{c1}$ and $\Delta\Omega_{c2}$. 
 The first (bulk laminar) regime, at $0< \Delta\Omega < \Delta\Omega_{\rm c1}$, in which $\Delta t$ and $\sigma$ stay unchanged and only $\theta$ grows linearly with increasing $\Delta \Omega$, corresponds to local vortex tangles near the top and bottom grid co-existing with an array of rectilinear vortex lines elsewhere in the bulk. 
With increasing forcing the density of the local tangle eventually becomes much larger than $L_0$, after which $\theta$ is expected to saturate below the  geometrical transparency of the grid assembly 0.64. 
Above $\Delta\Omega_{\rm c1}$, we do observe levelling-off at $\theta \approx 0.26$, which, according to Eq.\,\ref{theta}, corresponds to the local density $L\sim 10^4$\,cm$^{-2}$ (while $L_0 = 3\times 10^3$\,cm$^{-2}$). 

In the second regime, at $\Delta\Omega > \Delta\Omega_{\rm c2}$, most ions arrive at the collector much later than the peak injection at 0.20\,s, with a large spread of arrival times $\Delta t$ and with a larger lateral spread $\sigma$ than for the injection into an array of rectilinear vortex lines. 
This regime corresponds to a vortex tangle filling the whole container (bulk turbulence). The cross-over between the local and bulk turbulence takes place gradually between $\Delta\Omega_{\rm c1}$ and $\Delta\Omega_{\rm c2}$.
The first critical amplitude, $\Delta\Omega_{\rm c1}$, is set by the comparison of local $\ell$ near the collector grid with the grid period $s$. The second one $\Delta\Omega_{\rm c2}$, corresponds to $\Delta t$ and $\sigma$ becoming clearly greater than $\Delta t_0$ and $\sigma_0$ set by the injection's intensity and rotation rate $\Omega_0$.
 For the shown example of $\Omega_0 = 1.5$\,rad/s and $\omega = 1.5$\,rad/s, $\Delta\Omega_{c1} \approx 0.01$\,rad/s and $\Delta\Omega_{c2} \approx 0.03$\,rad/s; however, for $\Omega_0 = 1.5$\,rad/s and $\omega = 0.21$\,rad/s, $\Delta\Omega_{c1} \approx 0.04$\,rad/s and $\Delta\Omega_{c2} \approx 0.12$\,rad/s. 
For all studied frequencies $\omega$ and $\Omega_0$, we found that $\Delta\Omega_{\rm c2} \approx 3 \Delta\Omega_{\rm c1}$.   
The saturated value of $\theta$ was always within $0.25 \pm 0.1$. With the total injected charge $Q_{\rm tot} \equiv Q_{\rm coll} + Q_{\rm grid} + Q_{\rm plate}  = 23$\,pC and rotation rate $\Omega_0 = 1.5$\,rad/s, the low-forcing values were $\Delta t_0 = 0.21$\,s, $\sigma_0 = 0.29$. With decreasing $\Omega_0$, 
$\theta_0$ decreased while $\Delta t_0$ and $\sigma_0$ increased. 

 \begin{figure}
\centerline{\includegraphics[width=7.5cm]{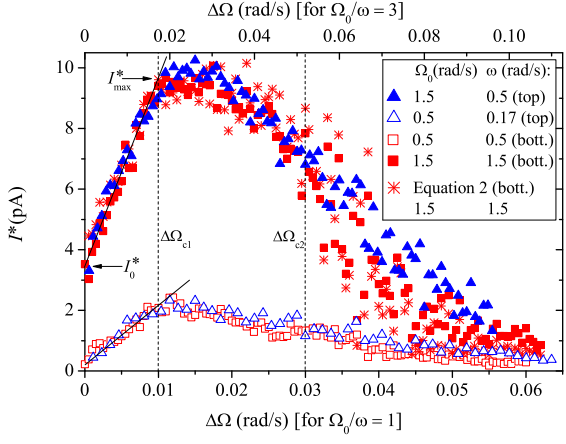}}
\caption{(color online) 
The amplitude of the collector current peak $I^*$ versus $\Delta\Omega$ for different $\Omega_0$ and $\omega$. The vertical lines indicate the critical amplitudes $\Delta\Omega_{\rm c1}$ and $\Delta\Omega_{\rm c2}$. The scale of the top horizontal axis (for $\Omega_0/\omega = 3$) is different from that of the bottom axis (for $\Omega_0/\omega = 1$) by the factor of 1.73. The solid lines guide the eye through the linear part. The values of $I^*(\Delta \Omega)$ for $\Omega_0 = \omega = 1.5$\,rad/s, calculated using Eq.\,\ref{Icoll} and data from Fig.\,\ref{FigTranspSpreadWidth}, are shown by asterisks. 
}
\label{FigPulseAmplScaled}
\end{figure}

To investigate the observed regimes at different $\Omega_0$ and $\omega$, we monitored the amplitude of the collector peak $I^*$, which is related (asterisks in Fig.\,\ref{FigPulseAmplScaled}) to the quantities $\theta$, $\sigma$ and $\Delta t$ through
\begin{equation}
	 I^* \approx 0.46Q_{\rm tot}\theta (1-\sigma)\Delta t ^{-1},
 \label{Icoll}
\end{equation} 
where 0.46 is an empirical factor. 
At small $\Delta\Omega < \Delta\Omega_{\rm c1}$, $I^*(\Delta\Omega)$ grows linearly, $I^* = I^*_0 + g_0 \Delta\Omega$, due to the growing $\theta (\Delta\Omega)$. At larger $\Delta\Omega > \Delta\Omega_{\rm c2}$, $I^*(\Delta\Omega)$ decreases chiefly because of the increasing $\Delta t (\Delta\Omega)$. 
As the maximum, $I^*_{\rm max}$, is just above $\Delta\Omega_{\rm c1}$, the gradient of the linear part of $I^* (\Delta\Omega)$, $g_0$, is inversly proportional to $\Delta\Omega_{\rm c1}$,
\begin{equation}
\Delta\Omega_{\rm c1} \approx g_0^{-1} \Delta I^*_{\rm max}  ,
\label{DWc1}	
\end{equation}
 where $\Delta I^*_{\rm max}(\Omega_0) \equiv I^*_{\rm max} - I^*_0$  is independent of $\omega$. 

\begin{figure}
\centerline{\includegraphics[width=7.5cm]{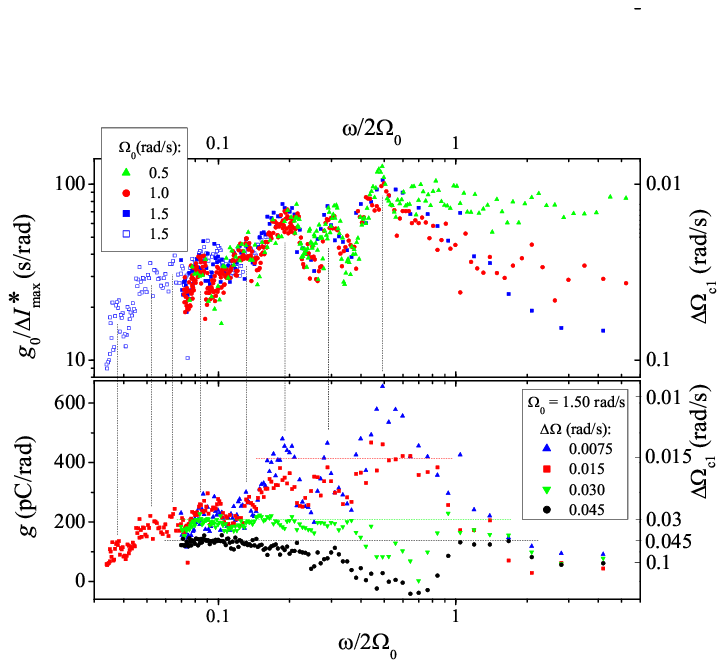}}
\caption{(color online) 
Top panel: the ratio of the gradient $g_0 \equiv (I^*(\Delta\Omega)-I^*_0)/\Delta\Omega$ of the linear part of $I^*(\Delta\Omega)$ to $\Delta I^*_{\rm max}(\Omega_0)$ versus the reduced forcing frequency $\omega/2\Omega_0$ for three rates of rotation $\Omega_0$. $g_0$ was measured at amplitude $\Delta \Omega = 0.0075$\,rad/s (except for open blue squares, for which $\Delta \Omega = 0.015$\,rad/s). $\Delta I^*_{\rm max}(0.5{\rm \,rad/s})=1.95$\,pA, $\Delta I^*_{\rm max}(1.0{\rm \,rad/s})=5.0$\,pA, $\Delta I^*_{\rm max}(1.5{\rm \,rad/s})=6.2$\,pA. The right axis shows the corresponding critical amplitudes $\Delta \Omega_{\rm c1}$ from Eq.\,\ref{DWc1}. Resonances are indicated by vertical lines.
Bottom panel: $g \equiv (I^*(\Delta\Omega)-I^*_0)/\Delta\Omega$ versus $\omega/2\Omega_0$ for $\Omega_0 = 1.50$\,rad/s and four different values of $\Delta\Omega$. The right axis shows the  amplitudes $\Delta \Omega_{\rm c1} = 6.2{\rm \,pA}/g$ for $\Omega_0 =$\,1.5 rad/s. The horizontal lines indicate the values of $\Delta\Omega_{\rm c1}(g)=\Delta\Omega$ beyond which the linear growth $I^*(\Delta \Omega)$ (i.\,e. $g=g_0$) breaks down; from top to bottom, $\Delta\Omega$: 0.015, 0.030, 0.045\,rad/s (the colors of lines and symbols for the same $\Delta\Omega$ are the same).}
\label{FigInertialWaves}
\end{figure}

In Fig.\,\ref{FigInertialWaves} (top panel), we plot the dependences on the forcing frequency $\omega$ of the ratio $g_0 /\Delta I^*_{\rm max}$ for three rates of rotation $\Omega_0$.  
There are several broad peaks at frequencies $\omega$ that are not fixed but proportional to the rotation rate $\Omega_0$. This is substantiated by plotting versus $\omega/2\Omega_0$: at $\omega/2\Omega_0 \leq 0.6$, all three datasets collapse on the unique function $\Delta\Omega_{\rm c1}^{-1}(\omega/2\Omega_0)$. 

 Resonances of inertial waves are indeed expected at certain $\omega/2\Omega_0 \leq 1$ \cite{Bewley2007}. These were calculated for an ideal liquid subject to non-slip boundary conditions in a rotating cube \cite{Maas2003} but generally disagree with the positions of our peaks.
 Whether this disagreement is because of different boundary conditions of turbulent superfluid in the $T=0$ limit or because our cell is not truly cubic (gaps between square electrodes at the edges of the cube and circular holes in the centres of four electrodes for injectors and collectors of ions) is an open question. 
 
The non-monotonic behaviour of $I^*(\Delta\Omega)$ at $\Delta\Omega > \Delta\Omega_{\rm c1}$ is presented in Fig.\,\ref{FigInertialWaves} (bottom panel).
Here we plot $g \equiv (I^*(\Delta\Omega)-I^*_0)/\Delta\Omega$.
When $\Delta\Omega > \Delta\Omega_{\rm c1}(\omega/2\Omega_0)$, $g$ no longer coincides with the gradient $g_0$ but falls below it. For each forcing amplitude $\Delta\Omega$, this limiting value of $g$, $g_0 = \Delta I^*_{\rm max} (\Omega_0)/\Delta\Omega$, is indicated by a horizontal line; these lines are in a good agreement with the observed plateaus (``chopped-off peaks'') in $g$. When $\Delta\Omega$ exceeds $\Delta\Omega_{\rm c1}(\omega/2\Omega_0)$, the peaks in $g_0(\omega/2\Omega_0)$ correspond to troughs in $g(\omega/2\Omega_0)$. 
 
We found that not only $\Delta\Omega_{\rm c1}(\omega,\Omega_0)$ stays the same if the ratio $\omega/\Omega_0$ is kept constant, but all $I^*(\Delta\Omega)$ for the same $\Omega_0$ collapse on a single curve if plotted versus $\Delta\Omega/\Delta\Omega_{\rm c1}(\omega, \Omega)$, Fig.\,\ref{FigPulseAmplScaled}. This reflects the fact (Eq.\,\ref{DWc1}) that all functions $\theta (\omega,\Omega_0)$, $\sigma (\omega,\Omega_0)$ and $\Delta t (\omega,\Omega_0)$ collapse on single curves when plotted this way (this was confirmed independently). In the investigated range of resonant frequencies, $0.05 \leq \omega/2\Omega_0 \leq 0.5$, the peak (on-resonance) values of $\Delta\Omega_{\rm c1}$ seem to scale as $\propto \omega^{-1/2}$, that can be summarized: $\Delta\Omega_{\rm c1} \sim 0.01(\omega/\Omega_0)^{-1/2}$\,rad/s. 
At troughs, although never farther away from a resonance than one linewidth, $\Delta\Omega_{\rm c1} \approx 0.04 \pm 0.01$\,rad/s is roughly independent of the forcing frequency. 
In the limit of high-frequency forcing $\omega/2\Omega_0 > 1$, where inertial waves cannot propagate, all $\Delta\Omega_{\rm c1}$ keep increasing in the manner perhaps compatible with the common scaling law $\Delta\Omega_{\rm c1} \propto (\omega/\Omega_0)^{1/2}$ \cite{Risto}, but eventually saturate at different values seemingly proportional to $\Omega_0$. At the highest frequency studied $\omega = 10$\,\/rad/s (at $\Omega_0 = 1.5$\,rad/s), $\Delta\Omega_{\rm c1}$ was about 10 times greater than that at the strongest inertial wave resonance (at $\omega/2\Omega_0 = 0.49$\,rad/s).  

 \begin{figure}
\centerline{\includegraphics[width=7.5cm]{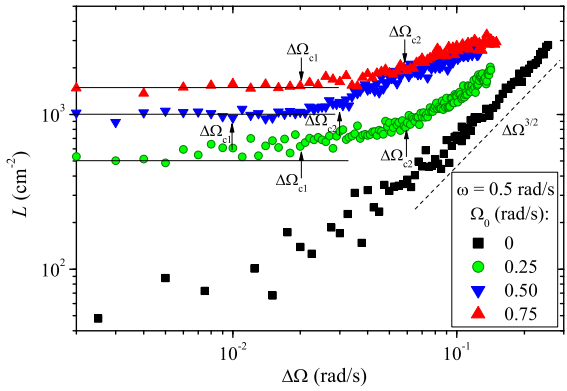}}
\caption{(color online) 
Mean values of vortex line density $L$, measured in the transverse direction, versus forcing amplitude $\Delta\Omega$ for various $\Omega_0$. Solid lines indicate the no-forcing limit $L_0 = 2\Omega_0/\kappa$. The arrows indicate the values of $\Delta\Omega_{\rm c1}$ and $\Delta\Omega_{\rm c2} = 3 \Delta\Omega_{\rm c1}$, where  $\Delta\Omega_{\rm c1} = \Delta I^*/g_0$ for particular values of $\omega/2\Omega_0$ were calculated using the data for $\Omega_0 = 1.5$\,rad/s: $\Delta I^*=6.2$\,pA and (see Fig.\,\ref{FigInertialWaves}, bottom panel) $g_0 = 300$\,pC/rad for $\omega/2\Omega_0 = 1$, $g_0 = 600$\,pC/rad for $\omega/2\Omega_0 = 0.5$ and $g_0 = 300$\,pC/rad for $\omega/2\Omega_0 = 0.33$.
}
\label{FiglogL}
\end{figure}

To confirm that in a turbulent state the length of the vortex tangle $L$ exceeds $L_0$, we measured its mean value along a horizontal axis of the cube via scattering of charged vortex rings \cite{WalmsleyJLTP2009} of average radius $R=1.6$\,$\mu$m sent from the left injector to the right collector. In Fig.\,\ref{FiglogL}, one can see that at forcing amplitudes smaller than $\Delta\Omega_{c2}$ the vortex line length, measured far from the horizontal walls, is only slowly increasing from the equilibrium $L_0=2\Omega_0/\kappa$.
But at $\Delta\Omega > \Delta\Omega_{c2}$, $L(\Delta\Omega)$ $L$ starts to grow faster and rapidly becomes much greater than $L_0$. This supports our interpretation that, at $\Delta\Omega > \Delta\Omega_{c2}$, a vortex tangle of low polarization develops in the bulk.

In the limit of slow rotation, the dependence $L(\Delta\Omega)$ in the turbulent regime (Fig.\,\ref{FiglogL}) is steeper than at $\Delta\Omega/\Omega_0 \ll 1$ and seems to be closer to $L \propto \Delta\Omega^{3/2}$ (as for $\Omega_0=0$). The exponent $3/2$  
 is indeed expected if we assume that the fully-developed turbulence exerts the resistive torque $\propto \Delta\Omega^2$;
 then its time-averaged work $\propto \Delta\Omega^3$ should equate the rate of dissipation $\propto \nu'(\kappa L)^2$. The critical amplitude for sustainable QT at $\Omega_0=0$ is  small but increases seemingly proportionally to $\Omega_0$. 

With fast rotation ($\Omega_0 > \omega/2$), inertial waves are an inherent feature of the large-scale superfluid dynamics, thus greatly reducing the critical amplitude of forcing at strongest resonances.
Between the resonances, the critical amplitudes are roughly frequency-independent,  $\Delta\Omega_{\rm c1} \approx 0.04$\,rad/s,  either due to the dominance of other mechanisms of generating turbulence or complex interaction of inertial waves of different wavenumbers. With further increasing $\Omega_0$, as the on-resonance critical amplitudes increase as $\propto (\Omega_0/\omega)^{1/2}$, one can expect the inertial waves to become non-dominant again if the forcing frequency $\omega$ is kept constant. 

{\bf Summary.}
This is the first experimental studies of rotating QT in the $T=0$ limit. 
The new detection techniques exploited the quantized nature of vorticity and were sensitive to the density and degree of entanglement of vortex lines separately in the bulk and near the container walls. Within our resolution, no finite critical amplitude of forcing was required to sustain either bulk turbulence in a non-rotating container or boundary turbulence near the walls of rotating container. 
In rotation, with increasing the amplitude of the AC-rotational forcing, we observed a gradual growth of vortex tangles near the horizontal walls followed by a transformation, around the finite critical amplitude of forcing $\Delta\Omega_{\rm c2}$, of an array of rectilinear vortex lines into a developed bulk vortex tangle. 
Inertial waves, observed at frequencies $\omega/2\Omega_0$ between 0.03 and 0.5, help generate turbulence: on-resonance, the critical amplitude $\Delta\Omega_{\rm c2} (\omega/2\Omega_0)$ can be several times smaller than when off-resonance. 

This work was supported by the
Engineering and Physical Sciences Research Council [grant numbers GR/R94855, EP/H04762X and EP/I003738].

\end{document}